# Severity of light pollution and its multifaceted impact: A mini review on the dark side of light

Pranjal Choudhary[1] and Sudhish Kumar[2]

[1]Jayshree Periwal Intentaional School, Jaipur, Rajasthan, India -302026

[2]Magnetism laboratory, Department of Physics, Mohanlal Sukhadia University, Udaipur, India -313002

## Abstract

There is a surging concern regarding the adverse effects of light pollution on human well-being. This manuscript aims to emphasise the deleterious effects of uncontrolled night-light exposure on the health and mood of individuals residing in a highly populated elite metropolitan society compared to those residing in remote villagesin a scattered manner. A comprehensive study has been undertaken on the influence of unnecessary and excessive illumination of light during late-night hours on individuals' day-to-day lives. Prolonged exposure to unwanted intense artificial lighting disrupts the secretion of melatonin, interferes with the circadian rhythm, and unusual sleep patterns. Such disturbances can lead to significant health issues, including insomnia, depression, cardiovascular diseases, as well as an increased risk of breast and prostate cancer. Findings emphasize the need for a more holistic understanding of how artificial light, in intensity and periodicity, affects the physical and mental health of humans and the ecological system in entirety, specifically in urban settings.

E-mail address: feninpranjal@gmail.com and skphy@mlsu.ac.in

## 1.0 Introduction

Over the years, design, fabrication, and the use of cheaper artificial lights has revolutionised the living conditions, quality of life, socio-economic cultural activities, and routine lives of humans (enhancing productivity, convenience, safety, and mobility); especially in night hours for individuals working on a 24/7 basis. However, in the past two decades unnecessary and excessive nocturnal artificial lighting have emerged as a global threat for the ecology, cleaner environment, animals, and human-life. Further, it is also illustrated that unwanted exposure to artificial lighting contributes to direct as well as indirect impacts on the health and mood of not only humans, but animals and insects as well. Strikingly, only limited literature is available on such an important topic. For instance, a recent systematic study demonstrated severe impact of light pollution on the environment, health, and behaviour of humans and animals.It was shown that reduction in artificial lights during night-time hours in urban environment is greatly beneficial to the human health and ecosystem[1]. In another study it was observed that the extensiveness of depressive symptoms along with suicidal behaviours was a result of excessive exposure to artificial light [2]. Although it is difficult to identify direct proofs of the detrimental impact of dim and unnecessary artificial light, there is growing evidence that even dim light exposure, especially at night may lead to not only dysthymic traitsbut also have deleterious effects on the women's body weight and activeness.

Moreover, too much use of artificial lighting also compromises the quality of nocturnal vision and interferes with the natural periodicity of the night sky. Further light pollution greatly disrupts the inherent circadian rhythms of humans. This paradigm-shift towards artificial lighting, while prodigious and revolutionary, exhibits the darker side of excessive use of lights. The substantial increment in the usage of artificial light and its applications must be addressed with careful-consideration and sustainable methods to make sure that it does not harm human health or nature. Inessential exposure to intense light sources above the recommended lighting threshold, menaces the human health, mental well-being, wildlife, and disrupts the ecological balance. The consequences of high levels of illumination in a high-rise, densely populated, upscale urban community have also been well documented in literature. Notably, excessive and uncontrolled use of artificial lighting often attributed by poor lighting design; thus, contributes heavily to light pollution. These factors generate substantial costs by negatively affecting several facets including wildlife, human well-being, mental health, astronomy, and energy conservation.

Nocturnal light exposure leads to eye strain, disrupt circadian physiology, and may also suppress the secretion of melatonin *(a hormone critical for sleep regulation)*. Furthermore, the retardation of sleep patterns can result in a long-term illness and other issues such as insomnia along with reduced sleep quality. These important health issues directly contribute to the severe development of chronic diseases. Moreover, such health-related problems contribute to delayed bedtimes, frequent awakenings, and obesity, which can negatively impactdigestion and overall well-being of a human being. In simple terms, light pollution is the consequence of the use of excessive and intense artificial lighting that is prevalent in urban environments, and thus poses significant challenges to human well-being, clean environment, and damages the natural world in several ways.

Although this study is a tiny effort on such a vast field however the in-depth analysis explores the extent of the issue of light pollution and its diverse harmful effects onindividuals, animals, and the ecosystem. Through a point to point survey,data was collected from a range of respondents which provided valuable insights into people's experiences, awareness, and perceptions regarding the impact of light pollution across multiple domains.

## 2.0 Methods

A mini-survey (by direct contact) was conducted among a diverse range of samples of an elite society of urban residents as well as scattered rural residents. In the survey, individuals belonging to different age groups and diverse backgrounds were well accommodated. A total of 300 participants partook a brief questionnaire with great sincerity and optimism, and thus rich data collection work was carried out and a comprehensive analysis was performed, on the direct impact of light pollution on the issues related to routine life of humans, behaviour of pets and domestic animals, and their negative effect on animals. The data collected were analysed using descriptive statistics, and the results were classified into four major categories: psychological, environmental, social, and mental impacts.

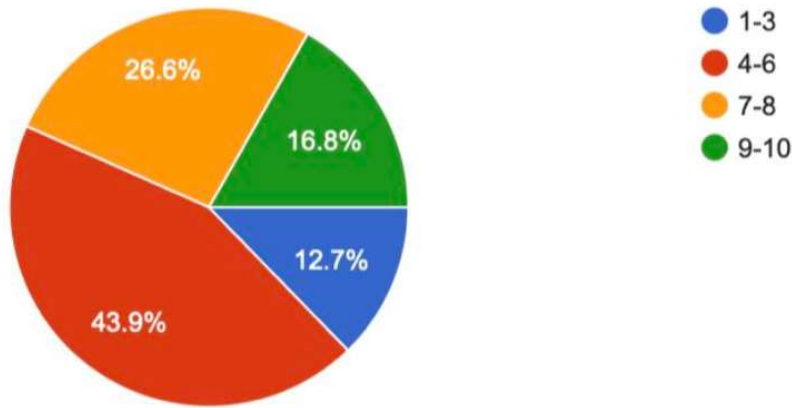

**Figure 1**: On a scale of 1 to 10, 300 people rated the level of lighting intensity,unwanted exposure to intense light source and its potential threatsto their sleep, health, and daily routines.

This survey was conducted in both, ahighly populated elite metropolitan society and in a remote village. It provided an understanding of and a closer look at the awareness about light pollution and its possible impacts on the physical as well as the mental health of not only human beings but animals and insects also.Surprisingly, people residing in the rural area were more understanding of the severity of the issue compared to the urbanites.

The geographical locations of the urban and rural sites under study are illustrated in the pictures given below.

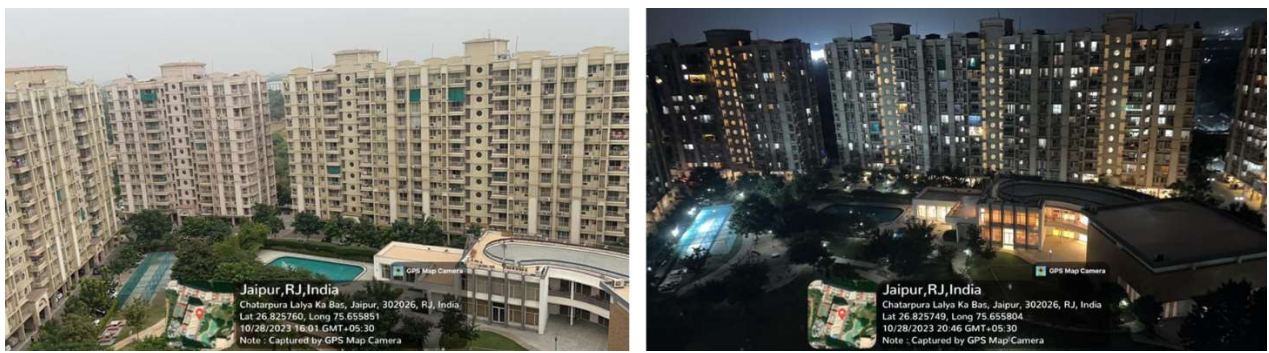

**Figure 2**: Image of the geographical location of the urban site during the day (left) and at night (right).

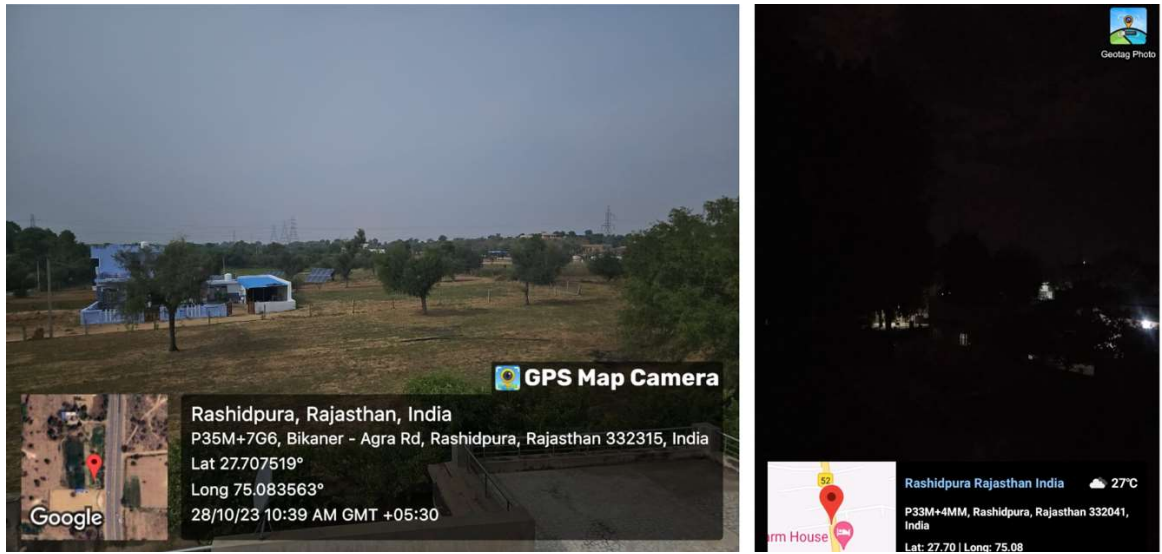

**Figure 3**: Image of the geographical location of the rural site during the day (left) and at night (right).

**3.0 Results and discussion**

The most lethal contributor to light pollution is the unnecessary and excessive exposure of artificial lights during late night hours (after 2200 hours) on the natural sleep-wake cycle of individuals. It is observed that street lights, and light pattern entering through the ventilation and glass-windows unnecessarily illuminate the bedrooms. Pie-chart in Figure 4 illustrates that 61.3% of respondents have been suffering fromdifficulty in falling asleep or staying asleep due to the unwanted exposure to intense lighting in the bedrooms. Because of the persistent exposure to bright as well as dim lights in bedrooms during sleeping hours at night directly interfere with the secretion of sleep hormone melatonin, which results in sleep deprived-ness and insomnia. Further, psychological toll of light pollution on sleep deprived-ness and poor sleep quality is a potential threat, which may lead to fatigue, irritability, and diminished health and wellness. In continuation, 60.1% of participants also reported a poor sleep pattern due to excessive lighting in their surroundings, such as bright lights in streets, roads, highways, and galleries in the residential apartments. Poor sleep quality is considerably associated with various health issues like impaired cognitive function and reduced overall well-being along with irritative behaviour.

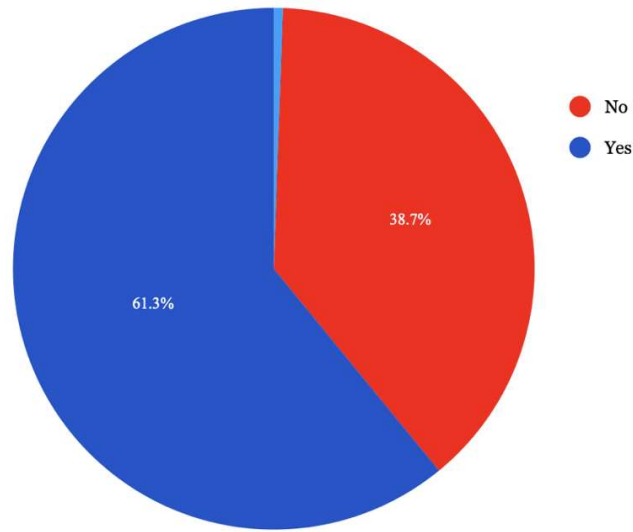

**Figure 4**: Personals experienced difficulty in falling asleep or staying asleep due to the unwanted presence of intense lighting in their bedrooms.

On the other hand, individuals residing in rural areas,in a scattered manner, have practicedand therefore become habitual to fall asleep at 2200 hrs and wake up at 0400 hrs to 0500 hrs, due to simplicity in their routine life and working meticulously during the daytime. This is why they have less exposure to light. Particularly, poor infrastructure; lower population density; and scattered houses also save them from unwanted exposure to artificial lights. Most of the villagers are less educated compared to urbanites but they value work ethics and importance of money.Thus, they also save a lot of electricity by timely utilisation of lights and this on-and-off mechanism is part of their routine life.

Besides above mentioned problemslight pollution also contributes to feelings of restlessness, anxiety, and irritation. Constant exposure to bright lights is bound to disrupt circadian rhythms and negatively affect mood regulation. Although not directly measured in the survey but the awareness of the detrimentalimpact associated with artificial light at night is significant, as over 64.7% of participants indicated a growing concern for the mental toll due to light pollution on the working personals life. Further, a substantial 57.8% (see Figure 5) of respondents complained about sleep deprivation due to the presence of unwanted, intense artificial lighting. In ambient conditions and surroundings, sleep deprivation has serious consequences on mental health, productivity, and immunity. Thus, making it a significant concern associated with threatening light pollution.

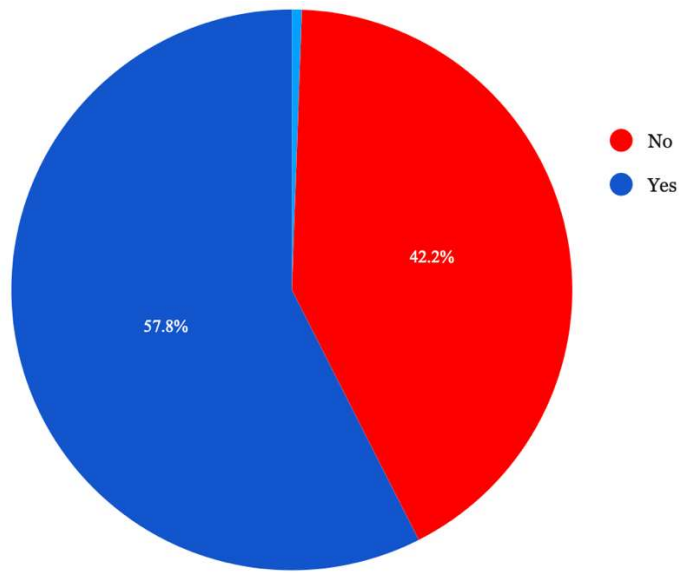

**Figure 5**: Personals experienced sleep deprivation due to the unwanted presence of intense artificial lighting in their surroundings.

Thanks to the simple lifestyle followed by most of the people residing in villages, they are unaffected from the above mentioned disorders. Moreover, light pollution badly disrupts the behaviour and natural rhythms of nocturnal animals and insects. It is noted that 75.7% of respondents believed that intense lighting hampers their ability to observe celestial objects and stars with naked eyes in the night sky. This disruption can have cascading effects on nocturnal ecosystems, affecting feeding patterns, mating behaviours, and predator-prey interactions. Further, 79% of participants acknowledged that excessive lighting significantly contributes to the wastage of power and financial resources. These findings not only leads to unnecessary energy consumption but also generates unnecessary greenhouse gas emissions, adding to the seriousthreat to the environment. During night, the bright lights can badly disorient migratory birds, leading to collisions with buildings or becoming easy targets for predators.

Additionally, artificial lighting considerably influences the reproductive success of some species and thus exacerbating biodiversity loss. Subsequently, a considerable majority, i.e., 73.4% of participants identified excessive lighting as an uninvited disturbance during the night-time hours. This sort-of disturbance is bound to disrupt social activities, hinder sleep, which is lethalfor the overall quality of life of urban dwellers. Contrary to popular belief, excessive lighting does not always enhance safety. Instead, it may create glare and shadows, hindering visibility and increase the risk of committing accidents and crime. Striking a balance between the safety features and responsible outdoor lighting is crucial for a happy and safer society.

Itwas also noted that 41% of respondents observed changes in the behaviour or sleep-activity patterns of domestic animals due to exposure to artificial lighting in night-hours. These findings underscore how light pollution affects the natural behaviours of animals, leading to alterations in their daily routines. Interestingly, 56.1% of participants were well aware of the potential negative effects of excessive artificial lighting on domestic animals' health and well-being. Disruptions in their sleep quality, hormone regulation, and overall physiological functioning can have adverse consequences for their overall health. 64.7% of respondents were well aware about the adverse effects associated with excessive exposure to artificial light, especially in late night hours. These observations clearly suggested a growing concern regarding the mental toll of light pollution on individuals' daily lives. Notably, 78% of participants faced negative impact of intense lighting on their eyes and vision which is resulting in the decrease in their visual acuity. These observations underscore the potential mental strain caused by light pollution.

In summary, light pollution is emerging as a pervasive and multifaceted issue with far-reaching impacts on human beings, cleaner environment, and the beauty of our earth. The detrimental effects of light pollution on the sleep quality, mental health, and welfarealong with disruption of nocturnal ecosystems and biodiversity, light pollution poses numerous challenges that demand immediate attention and intervention. Responsible outdoor lighting practices, public awareness campaigns, and policy interventions are essential in mitigating the harmful effects of light pollution and restoring the tranquillity of sleeping hours in nights and safer, cleaner atmosphere on the earth.

To further mitigate the serious problem of light pollution, necessary steps must be taken for addressing the psychological, environmental, social, and mental aspects of light pollution necessitates collaborative efforts among urban planners, policymakers, scientists, and citizens.

Adopting suitable technologies that minimises light wastage, implementing zoning regulations, and promoting community engagement can pave the way for a more sustainable and harmonious coexistence between urbanisation and nature. By recognising and addressing the complexities of light pollution, we can create a future where the beauty of starlit skies, the health of ecosystems, and the well-being of individuals will no longer be overshadowed by excessive artificial lighting.

## 4.0 Conclusion and Suggestions

Light pollution is emerging as a global threat. Excessive exposure to unnecessary lights leads to long-term illnesses and other issues such as insomnia, poor sleep quality, and uninvited development of chronic diseases. Moreover, light pollution also contributes to delayed bedtimes, frequent awakenings, obesity, anxiety, and depression.

Certain changes must be incorporated in order to reduce the impact of light pollution on our daily lives and creating a cleaner environment.Streetlights can be remodelled with a decreased tilt angles to limit direct light exposure andprohibiting light scatter. Use of white light sources must be minimised by discerning warmer alternatives to reduce disruption. Along with attenuation of indoor light pollution through innovative window design.In addition, installation of sensor-based lighting systems, that can be activated by motion or vehicle detection, so that lights can be switched on and off as per necessities.

Lastly, educating the public about the adverse effects of light pollution through awareness programs for making responsible lighting practices. Although this study is just a miniscule exploration of the impact of light pollution. We believe that work on the adverse effect of artificial lights would stimulate research in this emerging area.

## 5.0 Acknowledgements

# References

1. Zielinska-Dabkowska, K M et al. "Reducing Nighttime Light Exposure in the Urban Environment to Benefit Human Health and Society." Science (New York, N.Y.), vol. 380, no. 6650 (2023): 1130-1135. doi:10.1126/science.adg5277
2. Min, Jin-Young, and Min, Kyoung-Bok. "Outdoor Light at Night and the Prevalence of Depressive Symptoms and Suicidal Behaviors: A Cross-sectional Study in a Nationally Representative Sample of Korean Adults." Journal of Affective Disorders, vol. 227 (2018): 199-205. doi:10.1016/j.jad.2017.10.039.
3. Abay, Kibrom A, and Mulubrhan Amare. "Night light intensity and women's body weight: Evidence from Nigeria." Economics and Human Biology, vol. 31 (2018): 238-248. doi:10.1016/j.ehb.2018.09.001.
4. Aguilera-Benito, Patricia, et al. "Experimental Analysis of Passive Strategies in Houses with Glass Façades for the Use of Natural Light." Sustainability, vol. 13, no. 15, Aug. 2021, p. 8652. Crossref, https://doi.org/10.3390/su13158652.
5. Bista, A., Bista, D., Bhattarai, H., and Bhusal, P. "Study of the Impact of Lighting Intervention in the Historic and Touristic City of Nepal." 2023 IEEE Sustainable Smart Lighting World Conference & Expo (LS18), Mumbai, India, 2023, pp. 1-5. doi: 10.1109/LS1858153.2023.10170445.
6. Burt, Carolyn S et al. "The effects of light pollution on migratory animal behavior." Trends in Ecology & Evolution, vol. 38, no. 4 (2023): 355-368. doi:10.1016/j.tree.2022.12.006.
7. Cao, Miao, Ting Xu, and Daqiang Yin. "Understanding Light Pollution: Recent Advances on Its Health Threats and Regulations." Journal of Environmental Sciences, vol. 127, 2023, pp. 589-602. https://doi.org/10.1016/j.jes.2022.06.020.
8. Cissé, Yasmine M et al. "Depressive-like behavior is elevated among offspring of parents exposed to dim light at night prior to mating." Psychoneuroendocrinology, vol. 83 (2017): 182-186. doi:10.1016/j.psyneuen.2017.06.004.
9. Hölker, Franz, et al. "The Dark Side of Light: A Transdisciplinary Research Agenda for Light Pollution Policy." Ecology and Society, vol. 15, no. 13 (2010). doi:10.5751/ES-03685-150413.

10. Jägerbrand, Annika K, and Kamiel Spoelstra. "Effects of anthropogenic light on species and ecosystems." Science (New York, N.Y.), vol. 380, no. 6650 (2023): 1125-1130. doi:10.1126/science.adg3173.
11. Kaushik, Komal, Soumya Nair, and Arif Ahamad. "Studying Light Pollution as an Emerging Environmental Concern in India." Journal of Urban Management, vol. 11 (2022). doi:10.1016/j.jum.2022.05.012.
12. Lamphar, Héctor et al. "Light pollution as a factor in breast and prostate cancer." The Science of the Total Environment, vol. 806, Pt 4 (2022): 150918. doi:10.1016/j.scitotenv.2021.150918.
13. Mander, Susan, FakhrulAlam, Ruggiero Lovreglio, and Melanie Ooi. "How to Measure Light Pollution—A Systematic Review of Methods and Applications." Sustainable Cities and Society, vol. 92 (2023): 104465. doi:10.1016/j.scs.2023.104465.
14. Sung, Chan. "Examining the Effects of Vertical Outdoor Built Environment Characteristics on Indoor Light Pollution." Building and Environment, vol. 210 (2021): 108724.
15. Velasque, M et al. "Under the Influence of Light: How Light Pollution Disrupts Personality and Metabolism in Hermit Crabs." Environmental Pollution (Barking, Essex : 1987), vol. 316, Pt 2 (2023): 120594. doi:10.1016/j.envpol.2022.120594.
16. Wang, Tongyu et al. "Effects of Outdoor Artificial Light at Night on Human Health and Behavior: A Literature Review." Environmental Pollution (Barking, Essex : 1987), vol. 323 (2023): 121321. doi:10.1016/j.envpol.2023.121321.
17. Yan, Ziyan and Minghong Tan. "Changes in Light Pollution in the Pan-Third Pole's Protected Areas from 1992 to 2021." Ecological Informatics, vol. 75 (2023): 102016. https://doi.org/10.1016/j.ecoinf.2023.102016.
18. Zissis, Georges. "Sustainable Lighting and Light Pollution: A Critical Issue for the Present Generation, a Challenge to the Future." Sustainability, vol. 12, no. 11, June 2020, p. 4552. Crossref, https://doi.org/10.3390/su12114552.

# Appendix

1. Have you personally experienced difficulty falling asleep or staying asleep due to the presence of intense lighting in your vicinity?

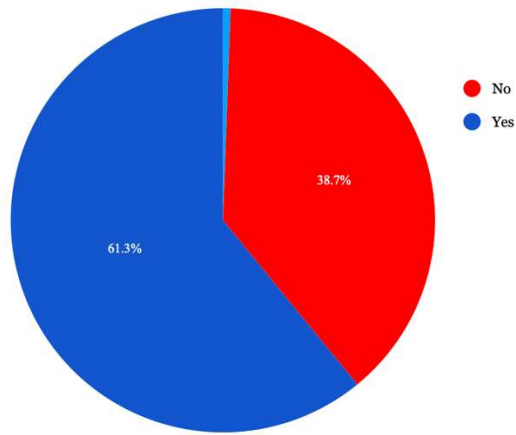

2. Have you experienced a decrease in the quality of your sleep due to excessive lighting in your surroundings, such as bright streetlights or gallery lights?

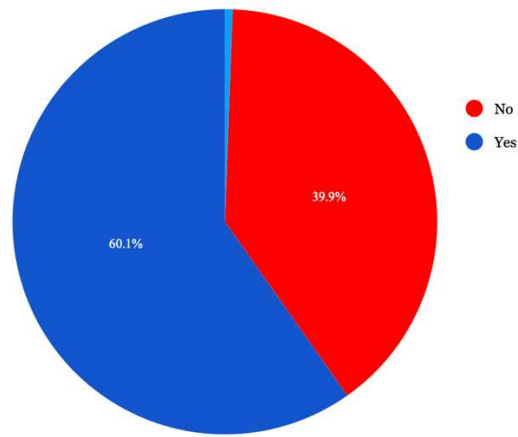

3. Have you experienced sleep deprivation specifically attributed to the presence of intense artificial lighting in your surroundings?

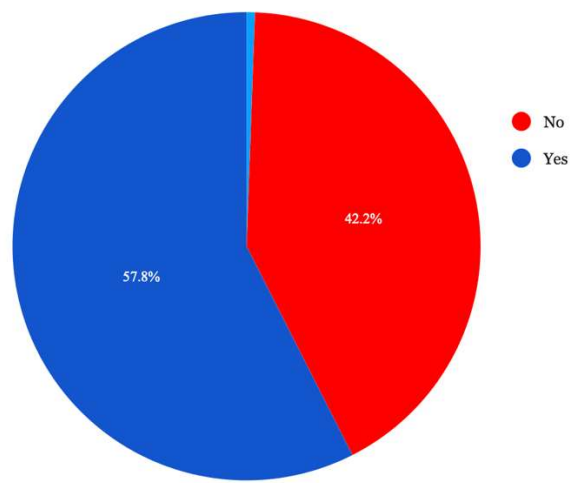

4. Are you aware about the adverse effects associated with excessive exposure to artificial light at night, including disruptions in sleep patterns and overall well-being?

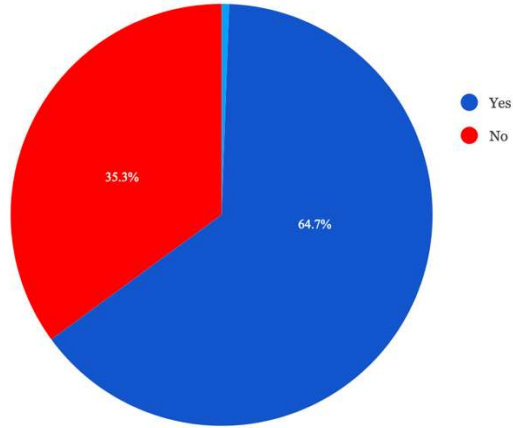

5. Do you believe that intense lighting has a negative impact on your eyes and vision, resulting in a decrease in visual acuity?

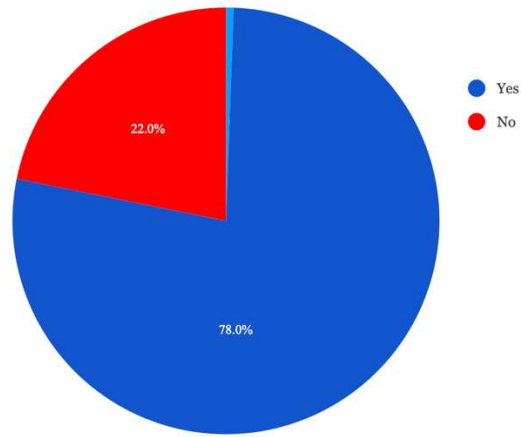

6. Do you think intense lighting hampers/impairs the ability to observe celestial objects and stars in the night sky?

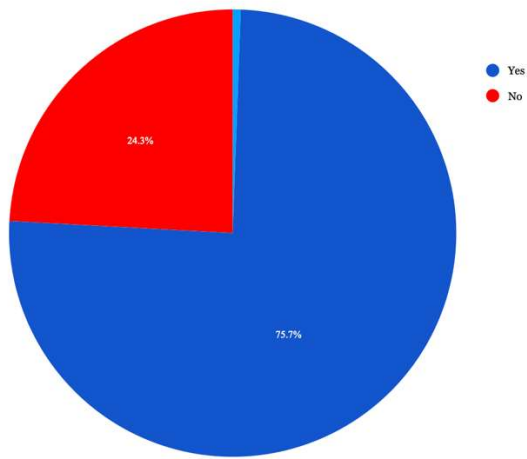

7. Do you believe that the issue of excessive lighting is significant in terms of causing unnecessary disturbance during nighttime hours?

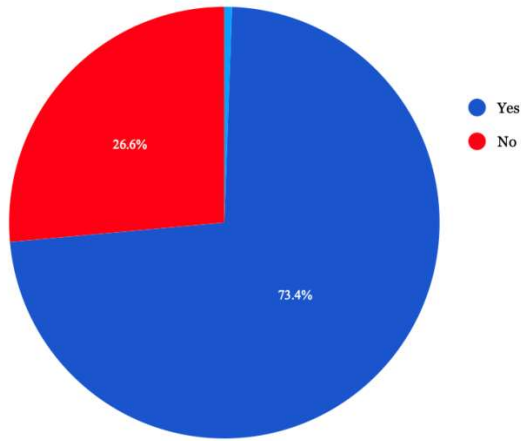

8. Which light source do you consider to be the best for vision among the following options:

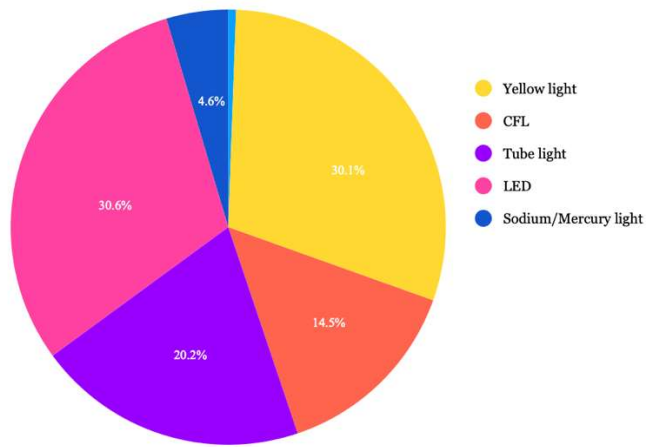

9. Have you observed the presence of small insects near white LEDs?

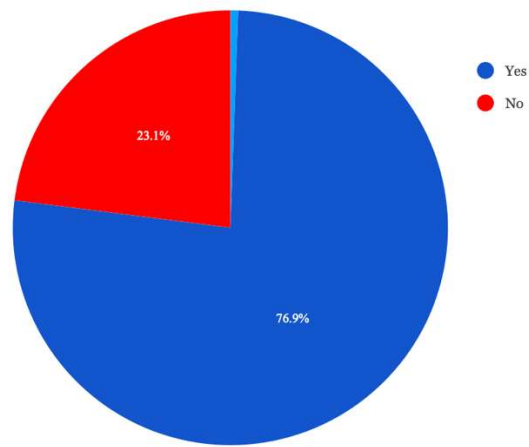

10. Have you noticed any changes in the behaviour or sleep/activity patterns of domestic animals in your household or neighbourhood that you believe are influenced by artificial lighting, such as streetlights or outdoor lighting?

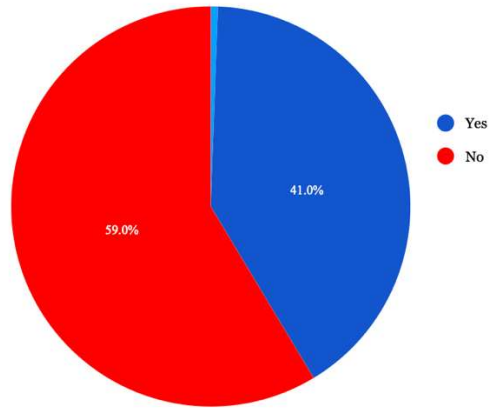

- Yes: 41.0%
- No: 59.0%

11. Are you aware of the potential negative effects of excessive artificial lighting on the health and well-being of domestic animals, including sleep quality, hormone regulation, and overall physiological functioning?

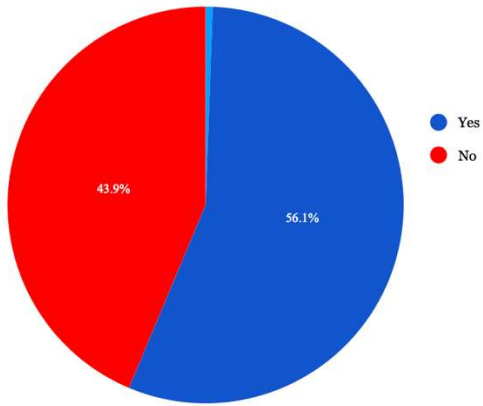

- Yes: 56.1%
- No: 43.9%

12. Have you observed any adverse effects on the sleep patterns or natural nocturnal behavior of domestic animals due to the presence of bright or intense lighting in their surroundings?

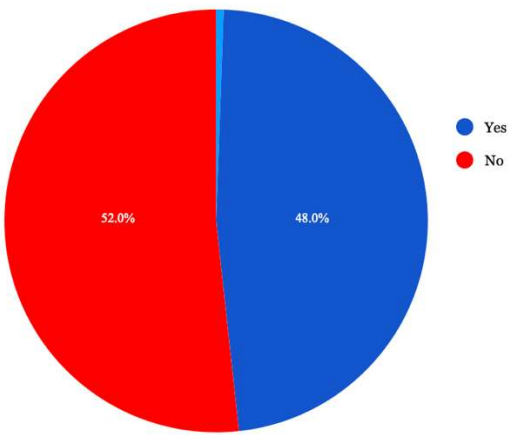

- Yes: 48.0%
- No: 52.0%

13. Are you convinced that excessive lighting contributes to the wastage of power and financial resources? Are you aware of the monetary cost incurred by your community or city due to excessive outdoor lighting in compounds?

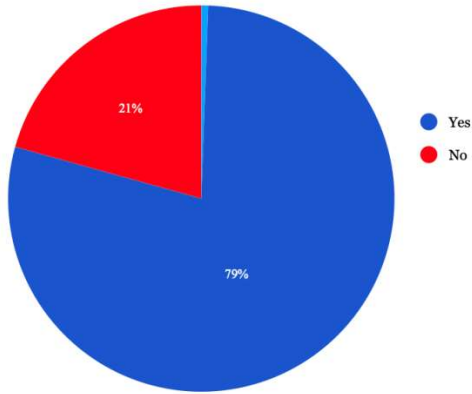

14. Do you believe that intense lighting is a significant problem in your neighbourhood? And that it is important for individuals and communities to take measures to minimise the effects of intense lighting?

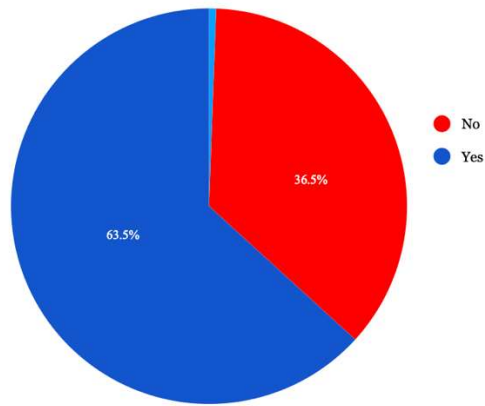